\documentclass[12pt]{iopart}

\usepackage{graphicx}
\usepackage{iopams}

\begin{document}

\newcommand{\beq}{\begin{equation}}
\newcommand{\eeq}{\end{equation}}
\newcommand{\beqarray}{\begin{eqnarray}}
\newcommand{\eeqarray}{\end{eqnarray}}
\newcommand{\bdisplay}{\begin{displaymath}}
\newcommand{\edisplay}{\end{displaymath}}
\newcommand{\lldots}{,\dots,}
\newcommand{\bk}{\bi{k}}
\newcommand{\bone}{\textbf 1}

\title{A parametrization of the baryon octet and decuplet masses}

\author{Phuoc Ha}

\address{
Department of Physics, Astronomy and Geosciences, Towson
University, Towson MD 21252, USA }

\ead{\mailto{pdha@towson.edu}}

\begin{abstract}

We construct a general parametrization of the baryon octet and
decuplet mass operators including the three-body terms using the
unit operator and the symmetry-breaking factors
$M^d=\textrm{diag}\,(0,1,0)$ and $M^s=\textrm{diag}\,(0,0,1)$ in
conjunction with the spin operators. Our parametrization has the
minimal number of operators needed to describe all the octet and
decuplet masses. Investigating the likely size of the three-body
terms, we find that contributions of the three-body hypercharge
splittings are comparable to those from the one- and two-body
isospin splittings and that contributions of the three-body
isospin splitting operators are very small. We prove that, in
dynamical calculations, one must go to three loops to get the
three-body terms. We also find that the suggested hierarchy of
sizes for terms in the most general expression for baryon masses
that involve multiple factors of $M^d$ and/or $M^s$ does not hold
strictly for dynamical calculations in heavy baryon chiral
perturbation theory:  terms of a given order in a meson loop
expansion may appear both with the expected factors of $M^d$ and
$M^s$, and with one factor more.

\end{abstract}


\maketitle

\section{Introduction}

The study of baryon masses is one of the fundamentally important
problems in nuclear and particle physics and has been of interest
for many years. Progress on the modern theory began with the
introduction of flavor SU(3) symmetry and of the electromagnetic
mass relation derived by Coleman and Glashow
\cite{Coleman-Glashow}. It was later found in the nonrelativistic
quark model \cite{Rubenstein1,Rubenstein2,Ishida,Gal-Scheck} that,
by including only two-body interactions among the quarks in the
baryons, there exist nine mass formulas (a.k.a. sum rules)
connecting the eighteen charge states of the baryon octet and
decuplet. One of them is the remarkably accurate Coleman-Glashow
relation. The baryon octet and decuplet masses and the sum rules
have been also investigated in various versions of chiral
perturbation theory
\cite{Jenkins_mass,BKM,Lebed-Luty,Bo-M,El-To,Jenkins-Lebed,Frink-Meissner}.
As remarked elsewhere, the chiral expansion gives a complete
parametrization of the static properties of the baryons when
carried to high enough order, but no dynamical information unless
the new couplings can be calculated in the underlying theory (see,
for example, the analysis of the magnetic moment problem in
\cite{DandH}).

It was Morpurgo who first constructed a general parametrization of
baryon masses for the case of hypercharge breaking. His general
parametrization method \cite{Morpurgo_param}, derived exactly from
the QCD Lagrangian employing only few general properties
\footnote{The general properties are listed in the first footnote
in \cite{DM-03}.}, expresses the possible mass operators in terms
of flavor-dependent terms proportional to powers of the
strange-quark projection operator $P^\lambda$ (denoted by $M^s$ in
our work), and nonrelativistic-appearing products of Pauli spin
operators $\bsigma$.  The results are completely general and
relativistic.  Morpurgo also considered the purely electromagnetic
contributions to the baryon masses and showed that the known sum
rules for mass splittings within isospin multiplets hold in the
absence of three-body terms \cite{Morpurgo_EM1}. Dillon and
Morpurgo later included the light quark mass terms in some of
their work. For example, they briefly discussed the effects of
$m_d-m_u$ on the Coleman-Glashow and other baryon mass formula in
\cite{Morpurgo_EM2} where the first order of the light quark mass
difference $m_d-m_u$ is considered.

As already mentioned, baryon masses have been studied extensively
in chiral perturbation theory (ChPT). See, for example,
\cite{Jenkins_mass,BKM,Lebed-Luty,Bo-M,El-To,Jenkins-Lebed,Frink-Meissner}
and the references in those papers.  The possible mass operators
are customarily expressed in terms of traces of products of baryon
effective fields and their derivatives using a matrix
representation of the fields, and mesonic corrections to the
initial operators are calculated in perturbation theory.  This
effective field theory approach is based on a well-defined
low-momentum approximation to QCD with a structure determined by
the general properties of QCD in the limit of small quark masses
and momenta \cite{Weinberg}, and gives a completely general
description of the low-energy properties of the baryons.

 In \cite{DHJ1,DHJ2,DH05}, we reconsidered the mass problem using
 the heavy-baryon version of chiral effective field theory
(HBChPT) \cite{Jenkins1}. We used a ``quark'' representation in
which the baryon effective fields are labeled using three flavor
indices $i,\,j,\,k$ and spin indices rather than the usual matrix
representation \footnote{The connection of this representation to
the usual effective-field methods was discussed in detail in
\cite{DHJ1,DHJ2}. The results in each case can be summarized in
terms of a set of effective interactions that have the appearance
of interactions between quarks in the familiar semirelativistic or
nonrelativistic quark models for the baryons
\cite{Brambilla,Carlson,Capstick}, but the corresponding matrix
elements can be translated back to expressions in terms of the
relativistic effective baryon fields.}, and calculated the
perturbative corrections in a meson-loop expansion.  We found that
this approach leads to a (relativistic) description of the
possible mass operators in HBChPT in terms of simple flavor and
spin operators which is identical to that developed by Morpurgo
\cite{Morpurgo_param}.  This equivalence is expected since HBChPT
retains the low-momentum structure of QCD, and the effective-field
results must be consistent with the general structure of operators
and matrix elements in QCD, and give a complete description of
those quantities at low energies where the effective field methods
apply. HBChPT also provides a formalism in which one can make
practical dynamical calculations perturbatively, an important
consideration in our earlier work.

In the present paper, we use Morpurgo's method
\cite{Morpurgo_param} to construct a general parametrization of
the baryon masses including three-body terms using the unit
operator and the symmetry-breaking factors,
$M^d=\textrm{diag}\,(0,1,0)$ and $M^s=\textrm{diag}\,(0,0,1)$, in
conjunction with the spin operators. Our general expression
includes all orders of the light quark mass difference $m_d-m_u$
which are not explicitly shown in the general parametrization for
baryon masses constructed by Dillon and Morpurgo. Using a Gram
matrix analysis, we establish a minimal  set of 18 independent
operators to describe 18 baryon octet and decuplet mass states.
The present work gives the overall setting without relying on
HBChPT.

Our objective in our earlier studies of mass splittings (and
moments) was to see the extent to which the observed splittings
(other than explicit quark mass terms) could be explained in terms
of dynamical interactions in HBChPT and the added electromagnetic
interactions \cite{DH05,Ha07}, rather than just parametrized.  Our
approach was therefore mainly calculational, restricted to one- or
two-loop corrections to the initial mass operators, and we did not
include an explicit analysis of the properties behind the
parametrization, or of the most general forms allowed.  We found
in that work that all the one-  and two-body operators were
generated in a meson loop expansion in HBChPT carried to two
loops, but no three-body operators were generated. We will show
here that one must go to three loops in HBChPT to get three-body
terms.  Our general demonstration of this and the examples of
nontrivial three-body terms are also new.

We will also investigate the likely sizes of the coefficients of
the various operators, particularly the three-body terms, in the
most general expression for baryon masses. We find that powers of
$M^d$ or $M^s$ do not correlate strictly with powers of the quark
masses $m_d-m_u$ or $m_s-m_u$ and, as a result, the suggested
hierarchy of sizes for terms in the most general expression for
baryon masses that involve multiple factors of $M^d$ and/or $M^s$
does not hold strictly for dynamical calculations in HBChPT.
Numerical calculations of the sum rules that give constraints on
the three-body terms indicate that contributions of the three-body
hypercharge splittings are comparable to those from the one- and
two-body isospin splittings. In addition, the calculations show
that contributions of the three-body isospin splitting operators
to baryon masses are very small. Ignoring terms associated with
the three-body isopsin splitting operators, we do phenomenological
fits to the experimental data using our general expression. Our
fits provide the values of the parameters that must be explained
in the loop expansion in ChPT or any other dynamical model. The
results of the fits also confirm that three-loop contributions
will necessarily play a role in explaining masses at the scale of
about 1 MeV.

Our parametrization has the minimal number of operators needed to
describe all the octet and decuplet masses and can be translated
back into HBChPT \cite{DHJ1,DHJ2}. Our general expression for
baryon masses is particularly useful to an analysis of the baryon
mass splittings due to simultaneous hypercharge-breaking and
isospin-breaking effects.

The paper is organized as follows. In Sec. \ref{GP-mass}, we
discuss how the general parametrization of the baryon octet and
decuplet masses is constructed.  A reduction of the general
expression to the independent one- and two body operators at
two-loop level and a general proof on the cancellation of the
three-body terms are shown in Sec. \ref{GP-twoloops}. Then, we
briefly discuss the hierarchy of sizes of the coefficients of the
various operators in the general expression for baryon masses,
investigate the likely size of the three-body terms, and do the
phenomenological fits to the experimental data in Sec.
\ref{hierarchy-and-fits}. Concluding remarks are presented in Sec.
\ref{conclusions}.

\section{General parametrization of the baryon octet and
decuplet masses} \label{GP-mass}

\subsection{Construction of the mass operators} \label{contruction}

First of all, we want to remind the reader that the operators that
appear, while nonrelativistic in appearance, are all that are
needed to describe the structure of matrix elements in exact QCD.

Before introducing our general parametrization of the baryon
masses, we need to show how to construct the possible mass
operators using the unit operator, $\bone$, and the
symmetry-breaking factors, $M^d=\textrm{diag}\,(0,1,0)$ and
$M^s=\textrm{diag}\,(0,0,1)$, in conjunction with the quark spin
operators $\bsigma$. Note that
$M^u=\textrm{diag}\,(1,0,0)=\bone-M^d-M^s$, so symmetry breaking
terms in $M^u$ can be transformed to $M^d$ and $M^s$. We are only
interested in the mass operators that can have nonzero matrix
elements in or between the octet and decuplet states.

If isospin breaking is ignored, employing the Morpurgo's general
parametrization method as discussed in \cite{Morpurgo_param}, one
can easily find that the following mass operators intervene in the
expression of the baryon masses:
\beqarray \label{hyperc-splittings}
\fl \bone \, , \, \, \sum_{i\not = j}\bsigma_i\cdot\bsigma_j \, , \nonumber  \\
\fl  \sum_iM^s_i \, , \,\, \sum_{i\not
=j}M^s_i\bsigma_i\cdot\bsigma_j \, , \,\, \sum_{i\not
=j}M^s_iM^s_j
\, , \,\, \sum_{i\not =j}M^s_iM^s_j\bsigma_i\cdot\bsigma_j \, ,  \\
\fl \sum_{i\not=j\not=k}M^s_i\bsigma_j\cdot\bsigma_k \, , \,\,
\sum_{i\not=j\not=k}M^s_iM^s_j\bsigma_j\cdot\bsigma_k \, , \,\,
\sum_{i\not=j\not=k}M^s_iM^s_jM^s_k \, , \,\,
\sum_{i\not=j\not=k}M^s_iM^s_jM^s_k\bsigma_j\cdot\bsigma_k \, ,
\nonumber \eeqarray
where $i,j,k \in u,d,s$ are flavor indices, $\bone$ and
$\sum_{i\not = j}\bsigma_i\cdot\bsigma_j$ are the flavor-symmetric
operators, the first line is the result for no flavor symmetry
breaking, and the second and third lines introduce the
hypercharge-breaking operators. Note that the operators carry the
``quark" labels of the effective fields that indicate the flavor
index on which they are to act. There are implied unit flavor
matrices for the quarks whose labels do not appear explicitly. For
example, $\sum_iM^s_i$ has the complete flavor structure
$(M^s_{i'i}\bone_{j'j}\bone_{k'k}
+\bone_{i'i}M^s_{j'j}\bone_{k'k}+\bone_{i'i}\bone_{j'j}M^s_{k'k}$).
In Eq. (\ref{hyperc-splittings}), terms in the first and second
lines are the one- and two-body operators and terms in the third
line are the three-body operators in the sense that they act
nontrivially on one, two, or three indices.

It is straightforward to generalize the result shown in Eq.
(\ref{hyperc-splittings}) to find the possible isospin splitting
operators - terms with one, two, or three factors of $M^d$. That
can be done by replacing in turn the matrix $M^s$ by $M^d$ at
every place it appears in a flavor splitting operator. For
example, the flavor splitting operators $\sum_iM^s_i$,
$\sum_{i\not =j}M^s_iM^s_j$, and
$\sum_{i\not=j\not=k}M^s_iM^s_j\bsigma_j\cdot\bsigma_k$ generate
the followings
\beqarray
\fl \sum_iM^s_i \rightarrow \sum_iM^d_i \, , \nonumber \\
\fl \sum_{i\not =j}M^s_iM^s_j \rightarrow \sum_{i\not
=j}M^d_iM^s_j \, , \, \, \sum_{i\not =j}M^s_iM^d_j \, , \, \,
\sum_{i\not =j}M^d_iM^d_j \, ,  \\
\fl \sum_{i\not=j\not=k}M^s_iM^s_j\bsigma_j\cdot\bsigma_k
\rightarrow \sum_{i\not=j\not=k}M^d_iM^s_j\bsigma_j\cdot\bsigma_k
\, , \, \, \sum_{i\not=j\not=k}M^s_iM^d_j\bsigma_j\cdot\bsigma_k
\, , \, \, \sum_{i\not=j\not=k}M^d_iM^d_j\bsigma_j\cdot\bsigma_k
\, \nonumber. \eeqarray
Since the indices $i$, $j$, and $k$ of the sums are dummy, one can
easily find that some of the generated operators are identical.
For example,
\beqarray \sum_{i\not =j}M^d_iM^s_j &=& \sum_{i\not =j}M^s_iM^d_j
\, , \, \,  \sum_{i\not =j}M^d_iM^s_j \bsigma_i\cdot\bsigma_j =
\sum_{i\not =j}M^s_iM^d_j\bsigma_i\cdot\bsigma_j \, , \nonumber
\\
\sum_{i\not =j\not=k}M^d_iM^s_jM^s_k &=& \sum_{i\not
=j\not=k}M^s_iM^d_jM^s_k=\sum_{i\not =j\not=k}M^s_iM^s_jM^d_k \, ,
\label{dumm-rels} \eeqarray
and so on. Keeping in mind Eq. (\ref{dumm-rels}), we find the
possible one- and two- body isospin splitting operators
\beqarray \label{group3} \fl \sum_iM^d_i \, , \,\, \sum_{i\not
=j}M^d_i\bsigma_i\cdot\bsigma_j \, , \,\, \sum_{i\not
=j}M^d_iM^s_j \, , \nonumber \\
\fl \sum_{i\not =j}M^d_iM^s_j\bsigma_i\cdot\bsigma_j \, , \, \,
\sum_{i\not =j}M^d_iM^d_j \, , \, \, \sum_{i\not
=j}M^d_iM^d_j\bsigma_i\cdot\bsigma_j \, , \eeqarray
and the possible three-body isospin splitting operators
\beqarray \fl \sum_{i\not=j\not=k}M^d_i\bsigma_j\cdot\bsigma_k \,
, \,\, \sum_{i\not=j\not=k}M^d_iM^s_j \bsigma_i\cdot\bsigma_k \, ,
\,\, \sum_{i\not=j\not=k}M^d_iM^s_j \bsigma_j\cdot\bsigma_k \, ,
\nonumber
\\
\fl \sum_{i\not=j\not=k}M^d_iM^s_jM^s_k \, , \,\,
\sum_{i\not=j\not=k}M^d_iM^s_jM^s_k \bsigma_i\cdot\bsigma_j \, ,
\,\, \sum_{i\not=j\not=k}M^d_iM^s_jM^s_k \bsigma_j\cdot\bsigma_k
\, , \,\, \nonumber
\\
\fl \sum_{i\not=j\not=k}M^d_iM^d_j \bsigma_j\cdot\bsigma_k \, ,
\,\, \sum_{i\not=j\not=k}M^d_iM^d_jM^s_k \, , \, \,
\sum_{i\not=j\not=k}M^d_iM^d_jM^s_k \bsigma_i\cdot\bsigma_j \, ,
\, \, \sum_{i\not=j\not=k}M^d_iM^d_jM^s_k \bsigma_i\cdot\bsigma_k
\, ,
\nonumber \\
\fl \sum_{i\not=j\not=k}M^d_iM^d_jM^d_k \, , \, \,
\sum_{i\not=j\not=k}M^d_iM^d_jM^d_k \bsigma_i\cdot\bsigma_j \, .
\label{group4} \eeqarray

We note that some isospin splitting operators listed in Eqs.
(\ref{group3}) and (\ref{group4}) (e.g., terms involving multiple
factors of $M^d$)  are not explicitly shown in the parametrization
constructed by Dillon and Morpurgo. However, it is worth pointing
out that, using the identities
\beq \label{s_q} Q = \frac{2}{3}\bone - M^d - M^s \, , \, \, Q^2 =
\frac{4}{9}\bone - \frac{1}{3}M^d - \frac{1}{3}M^s \, \eeq
for the charge operator $Q=\textrm{diag}\,(2/3,-1/3,-1/3)$, and
also the matrix relations $M^s_iM^d_i=M^d_iM^s_i=0$, $M^dM^d=M^d$,
and $M^sM^s=M^s$, the electromagnetic operators introduced by
Morpurgo in \cite{Morpurgo_EM1} can be reduced to sums of the
flavor-symmetric operators, the hypercharge splitting operators
shown in Eq. (\ref{hyperc-splittings}), and the isospin splitting
operators (except those with three factors of $M^d$) listed in
Eqs. (\ref{group3}) and (\ref{group4}), respectively.

\subsection{A general expression of the baryon masses}
\label{gen-exp}

Not all the splitting operators shown in Eqs.
(\ref{hyperc-splittings}), (\ref{group3}), and (\ref{group4})  are
independent. Using the identity $M^p_{i'j}M^p_{j'i} =
M^p_{i'i}M^p_{j'j}$, where $p=u,d,s$, and the exchange operator
$P_{ij}=(1+\bsigma_i\cdot\bsigma_j)/2$ from \cite{DHJ2} to
rearrange indices, we can easily show that
\beq M^p_iM^p_j=M^p_iM^p_j \bsigma_i\cdot\bsigma_j \, .
\label{propM1} \eeq
As a result, we obtain the following identities
\beqarray
\sum_{i\not =j}M^p_iM^p_j &=& \sum_{i\not=j}M^s_iM^s_j
\bsigma_i\cdot\bsigma_j \, , \nonumber \\
\sum_{i\not=j\not=k}M^p_iM^p_jM^q_k &=&
\sum_{i\not=j\not=k}M^p_iM^p_jM^q_k \bsigma_i\cdot\bsigma_j \, ,
\label{propM2} \eeqarray
where $p,q=u,d,s$.
Taking into account these identities, we find that the numbers of
operators in Eqs. (\ref{hyperc-splittings}), (\ref{group3}) and
(\ref{group4}) are reduced to 8, 5, and 9, respectively.
Hereafter, for the operators related by Eq. (\ref{propM2}), we
choose to work with those without spin operators.

We now have 22 operators (2 symmetric, 6 hypercharge splitting, 5
one- and two-body isospin splitting, and 9 three-body isospin
splitting) to describe a total of 18 baryon octet and decuplet
masses. Therefore, four operators must be redundant. To determine
what operators are independent, we make the tables of their
contributions to the baryon masses and then consider the
corresponding Gram matrix. For the eight flavor-symmetric
operators and the hypercharge splitting operators, their
contributions to the baryon masses are given in Table
\ref{sym-hyperc} where the three-body hypercharge splitting
operators are denoted as
\beq \label{3b-hop} \fl
s_1=\sum_{i\not=j\not=k}M^s_i\bsigma_j\cdot\bsigma_k \, , \, \,
s_2=\sum_{i\not=j\not=k}M^s_iM^s_j\bsigma_j\cdot\bsigma_k \, , \,
\, s_3=\sum_{i\not=j\not=k}M^s_iM^s_jM^s_k \, . \eeq
\begin{table}
\caption{Contributions of the flavor-symmetric operators and the
hypercharge splitting operators listed in Eq.
(\ref{hyperc-splittings}) to the baryon masses. For simplicity,
all the sums are suppressed. Contributions of the dependent
operators identified by Eq. (\ref {propM2}) are not shown. The
matrix elements for all particles in a hypercharge multiplet are
the same so that the table is actually $18\times 8$.}
\begin{indented}
\item[]
\begin{tabular}{@{}lccccccccc}
\br Baryon & $\bone$ & $\bsigma_i\cdot\bsigma_j$ & $M^s_i$ &
$M^s_i\bsigma_i\cdot\bsigma_j$ & $M^s_iM^s_j$ & $s_1$ &
$s_2$ & $s_3$ \\
\mr
$N$       & 1 & $-$6 & 0 & 0  & 0 & 0    & 0  & 0  \\
$\Lambda$ & 1 & $-$6 & 1 & 0  & 0 & $-$6 & 0  & 0 \\
$\Sigma$  & 1 & $-$6 & 1 & -4 & 0 & 2    & 0  & 0 \\
$\Xi$     & 1 & $-$6 & 2 & -2 & 2 & $-$8 & $-$4  & 0 \\
\mr
$\Delta$     & 1 & 6 & 0  & 0  & 0 & 0  & 0  & 0 \\
$\Sigma^{*}$ & 1 & 6 & 1  & 2  & 0 & 2  & 0  & 0 \\
$\Xi^{*}$    & 1 & 6 & 2  & 4  & 2 & 4  & 2  & 0 \\
$\Omega$     & 1 & 6 & 3  & 6  & 6 & 6  & 6  & 6 \\
\br
\end{tabular}
\label{sym-hyperc}
\end{indented}
\end{table}
We present in Table \ref{IB-terms} the contributions of the
thirteen isospin splitting operators where, for simplicity, the
three-body isospin splitting operators are labelled as
\beqarray \fl t_1=\sum_{i\not=j\not=k}M^d_iM^s_jM^s_k \, , \,\,
t_2=\sum_{i\not=j\not=k}M^d_iM^s_jM^s_k \bsigma_i\cdot\bsigma_j \,
, \,\, t_3=\sum_{i\not=j\not=k}M^d_iM^d_j \bsigma_j\cdot\bsigma_k
\, , \nonumber
\\
\fl t_4=\sum_{i\not=j\not=k}M^d_iM^d_jM^s_k \, , \, \,
t_5=\sum_{i\not=j\not=k}M^d_iM^d_jM^s_k \bsigma_i\cdot\bsigma_k \,
, \, \, t_6=\sum_{i\not=j\not=k}M^d_iM^d_jM^d_k  \, ,
\label{3-bop}
\\
\fl
t_7=\sum_{i\not=j\not=k}M^d_i\bsigma_j\cdot\bsigma_k \, , \,\,
t_8=\sum_{i\not=j\not=k}M^d_iM^s_j \bsigma_i\cdot\bsigma_k \, ,
\,\, t_9=\sum_{i\not=j\not=k}M^d_iM^s_j \bsigma_j\cdot\bsigma_k
\nonumber \, . \eeqarray
\begin{table}
\caption{Contributions of the isospin splitting operators listed
in Eqs. (\ref{group3}) and (\ref{group4}) to the baryon masses.
Here, for simplicity, we denote $c_1=\sum_iM^d_i$,
$c_2=\sum_{i\not =j}M^d_i\bsigma_i\cdot\bsigma_j$,
$c_3=\sum_{i\not =j}M^d_iM^s_j$, $c_4=\sum_{i\not
=j}M^d_iM^s_j\bsigma_i\cdot\bsigma_j$, and $c_5=\sum_{i\not
=j}M^d_iM^d_j$. Contributions of the dependent operators
identified by Eq. (\ref {propM2}) are not shown.}
\begin{indented}
\item[]
\begin{tabular}{@{}lcccccccccccccc}
\br Baryon & $c_1$ & $c_2$ & $c_3$ & $c_4$ & $c_5$ & $t_1$ & $t_2$
& $t_3$ & $t_4$ & $t_5$ & $t_6$ &
$t_7$ & $t_8$ & $t_9$ \\
\mr
$p$ & 1  & $-$4  & 0  & 0  & 0 & 0  & 0 & 0 & 0 & 0 & 0 & 2  & 0  & 0  \\
$n$ & 2  & $-$2  & 0  & 0  & 2 & 0  & 0 & $-$4 & 0 & 0 & 0 & $-$8 &  0  & 0  \\
$\Lambda$ & 1  & $-$3  & 1  & 0  & 0 & 0  & 0 & 0 & 0 & 0 & 0 & 0  & $-$3  & 0   \\
$\Sigma^+$ & 0 & 0  & 0  & 0  & 0 & 0 & 0  & 0  & 0  & 0 & 0 & 0 & 0 & 0 \\
$\Sigma^0$ & 1 & $-$1  & 1  & $-$2  & 0 & 0  & 0 & 0 & 0 & 0 & 0 & $-$4 &  1  & $-$2  \\
$\Sigma^-$ & 2  & $-$2  & 2  & $-$4  & 2 & 0  & 0 & $-$4 & 2 & $-$4 & 0 & $-$8 &  2  & $-$4  \\
$\Xi^0$ & 0 & 0  & 0  & 0  & 0 & 0 & 0  & 0  & 0  & 0 & 0 & 0 & 0 & 0 \\
$\Xi^-$ & 1  &  $-$4  & 2  & $-$4  & 0 & 2 & $-$4  & 0 & 0 & 0 & 0 & 2  &  $-$4  & 2  \\
\hline
$\Delta^{++}$ & 0 & 0  & 0  & 0  & 0 & 0 & 0  & 0  & 0  & 0 & 0  & 0  & 0  & 0 \\
$\Delta^+$ & 1  & 2  & 0  & 0  & 0  & 0  & 0 & 0  & 0  & 0  & 0 & 2 & 0 & 0  \\
$\Delta^0$& 2 & 4  & 0  & 0  & 2  & 0  & 0 & 2  & 0  & 0  & 0 &  4 & 0 & 0  \\
$\Delta^-$ & 3  & 6  & 0  & 0  & 6  & 0  & 0 & 6  & 0  & 0  & 6 &  6 & 0 & 0  \\
$\Sigma^{*+}$ & 0 & 0  & 0  & 0  & 0 & 0 & 0  & 0  & 0  & 0 & 0  & 0  & 0  & 0 \\
$\Sigma^{*0}$ & 1 & 2  & 1  & 1  & 0  & 0  & 0 & 0  & 0  & 0  & 0 & 2 & 1 & 1  \\
$\Sigma^{*-}$ & 2  & 4  & 2 & 2  & 2  & 0  & 0  & 2 & 2 & 2  & 0 & 4 & 2  & 2  \\
$\Xi^{*0}$ & 0 & 0  & 0  & 0  & 0 & 0 & 0  & 0  & 0  & 0 & 0  & 0  & 0  & 0 \\
$\Xi^{*-}$ & 1  &  2  & 2  & 2  & 0 & 2  & 2 & 0  & 0  & 0  & 0 & 2  &  2  & 2  \\
$\Omega^-$ & 0 & 0  & 0  & 0  & 0 & 0 & 0  & 0  & 0  & 0 & 0  & 0  & 0  & 0 \\
\br
\end{tabular}
\label{IB-terms}
\end{indented}
\end{table}
We then consider the $22\times 22$ Gram matrix
$\mathcal{M}_{\Gamma}=\mathcal{M}^T\mathcal{M}$ associated with
the $18\times 22$ matrix $\mathcal{M}$ defined by joining the
$18\times 8$ matrix in Table \ref{sym-hyperc} (extended to all
states in each hypercharge multiplet) and the $18\times 14$ matrix
in Table \ref{IB-terms}. $\mathcal{M}_{\Gamma}$ has a vanishing
determinant and four zero eigenvalues. This indicates that there
are four relations among the 22 operators. Applying the conditions
for linear dependence of the mass operators, $\mathcal{M}_{\Gamma}
\chi = 0$ for the eigenvectors $\chi$ with zero eigenvalues, we
find the four following relations:
\beqarray \label{4rel1} \fl 0 = -6 \, \bone + \sum_{i\not
=j}\bsigma_i\cdot\bsigma_j
+ 4\sum_iM^s_i -2 \sum_{i\not=j}M^s_i\bsigma_i\cdot\bsigma_j \nonumber \\
+ 4\sum_iM^d_i - 2\sum_{i\not =j}M^d_i\bsigma_i\cdot\bsigma_j -
2\sum_{i\not =j}M^d_iM^s_j + 2 \sum_{i\not
=j}M^d_iM^s_j\bsigma_i\cdot\bsigma_j \, , \eeqarray
\beqarray
 \label{4rel2} \fl 0 =
-6 \, \bone + \sum_{i\not=j}\bsigma_i\cdot\bsigma_j + 6\sum_iM^s_i
-2 \sum_{i\not =j}M^s_i\bsigma_i\cdot\bsigma_j  -2 \sum_{i\not
=j}M^s_iM^s_j - s_1 +
2s_2 \nonumber \\
+ 4\sum_iM^d_i - 2\sum_{i\not =j}M^d_i\bsigma_i\cdot\bsigma_j -
4\sum_{i\not =j}M^d_iM^s_j + 2t_8 + 2t_9  \, , \eeqarray
\beqarray
 \label{4rel3}
\fl 0 = -6 \, \bone + \sum_{i\not =j}\bsigma_i\cdot\bsigma_j  +
6\sum_iM^s_i -2 \sum_{i\not =j}M^s_i\bsigma_i\cdot\bsigma_j -2
\sum_{i\not =j}M^s_iM^s_j -s_1 + 2s_2 \nonumber \\ + 2\sum_iM^d_i
- 2\sum_{i\not =j}M^d_i\bsigma_i\cdot\bsigma_j - 2\sum_{i\not
=j}M^d_iM^s_j + 2\sum_{i\not =j}M^d_iM^d_j  \nonumber
\\ + t_7+2t_8
- 2t_3 -2t_4 + 2t_5  \, , \eeqarray \beqarray
 \label{4rel4}
\fl 0 = 2\sum_iM^s_i + 2\sum_{i\not =j}M^s_iM^s_j + s_1 - 2s_2 +
2\sum_{i\not =j}M^d_iM^s_j - 2t_8 - 2t_1 + 2t_2  \, . \eeqarray

Using these relations, we can select 4 operators to be dependent.
The chosen operators are $\sum_{i\not
=j}M^d_iM^s_j\bsigma_i\cdot\bsigma_j$, $t_7$, $t_8$, and $t_9$.
Note that the chosen two-body operator is related in Eq.
(\ref{4rel1}) only to other one- and two-body operators and does
not change the values of any three-body operators. On the other
hand, it follows from the remaining equations that the matrix
elements of $t_7$, $t_8$, and $t_9$ can be absorbed in the matrix
elements of one- and two-body operators and the remaining
three-body operators  $(s_1 - 2s_2)$, $t_1$ to $t_5$. No one- or
two-body operators are mixed into the three-body operators, so the
expected smallness of the remaining three-body terms should be
maintained.

We now have a following set of 18 independent mass operators
\beqarray \fl \mbox{\underline{Flavor-symmetric}} & \bone \, , \,
\,
\sum_{i\not = j}\bsigma_i\cdot\bsigma_j \, , \label{symmetric}  \\
\fl \mbox{\underline{Hypercharge splitting}} & \nonumber  \\
\fl \mbox{One and two-body:} & \sum_iM^s_i \, , \,\, \sum_{i\not
=j}M^s_i\bsigma_i\cdot\bsigma_j
\, , \,\, \sum_{i\not =j}M^s_iM^s_j \, , \nonumber \\
\fl \mbox{Three-body:} &
\sum_{i\not=j\not=k}M^s_i\bsigma_j\cdot\bsigma_k \, , \,\,
\sum_{i\not=j\not=k}M^s_iM^s_j\bsigma_j\cdot\bsigma_k \, ,
\,\,\sum_{i\not=j\not=k}M^s_iM^s_jM^s_k \, .  \label{group1p}\\
\fl \mbox{\underline{Isospin splitting}} & \nonumber  \\
\fl \mbox{One and two-body:}& \sum_iM^d_i \, , \,\, \sum_{i\not
=j}M^d_i\bsigma_i\cdot\bsigma_j \, \, , \, \, \sum_{i\not
=j}M^d_iM^d_j  \, , \nonumber \\
\fl \mbox{Three-body:} & \sum_{i\not=j\not=k}M^d_iM^d_j
\bsigma_j\cdot\bsigma_k \, , \,\,
\sum_{i\not=j\not=k}M^d_iM^d_jM^d_k
\, , \label{group2p} \\
\fl \mbox{\underline{Mixed splitting}} & \nonumber  \\
\fl \mbox{Two-body:}& \sum_{i\not =j}M^d_iM^s_j  \, , \nonumber
\\
\fl \mbox{Three-body:} &\sum_{i\not=j\not=k}M^d_iM^s_jM^s_k \, ,
\,\, \sum_{i\not=j\not=k}M^d_iM^s_jM^s_k \bsigma_i\cdot\bsigma_j
\, , \,\,
\sum_{i\not=j\not=k}M^d_iM^d_jM^s_k \, , \nonumber  \\
& \sum_{i\not=j\not=k}M^d_iM^d_jM^s_k \bsigma_i\cdot\bsigma_k \, .
\label{group3p} \eeqarray
Note that the mixed splitting operators affect both hypercharge
and isospin splittings and involve at least one factor each of
$M^s$ and $M^d$.

We can now construct a general expression for baryon masses. Using
the flavor-symmetric operators and other independent mass
operators given in Eqs. (\ref{symmetric}) - (\ref{group3p}), we
write the most general expression for baryon masses as
\beq \label{mass-gen} \mathcal{H}_B = m_0 \bone + A \sum_{i\not =
j}\bsigma_i\cdot\bsigma_j + \mathcal{H}_{Inter} +
\mathcal{H}_{Intra} \, , \eeq
where $\mathcal{H}_{Inter}$ and $\mathcal{H}_{Intra}$ are the
general expressions for the intermultiplet splittings and
intramultiplet splittings, respectively. Namely,
\beq \label{Hs} \fl \mathcal{H}_{Inter} =B \sum_iM^s_i + C
\sum_{i\not =j}M^s_i\bsigma_i\cdot\bsigma_j + D \sum_{i\not
=j}M^s_iM^s_j
 + a s_1 + b s_2 + c s_3 \, , \eeq
and
\beq \label{Hib} \fl \mathcal{H}_{Intra} = A' \sum_iM^d_i + B'
\sum_{i\not =j}M^d_i\bsigma_i\cdot\bsigma_j + C' \sum_{i\not
=j}M^d_iM^s_j + D' \sum_{i\not =j}M^d_iM^d_j + \sum_{i=1}^6 d_i
t_i  \, . \eeq
Here $m_0$, $A$, $B$, $C$, $D$, $A'$, $B'$, $C'$, $D'$, $a$, $b$,
$c$, and $d_i$ $(i=1,..., 6)$ are the parameters.

\section{A parametrization of baryon masses to two loops}
\label{GP-twoloops}

\subsection{Generation of the one- and two-body operators}

In \cite{DHJ2}, we analyzed the structure of meson loop
corrections to the $O(m_s)$ expressions for the baryon masses. Our
approach was based on the standard chiral Lagrangian of the
heavy-baryon chiral effective field theory \cite{Jenkins1}.
However, we used a spin- and flavor-index or ``quark''
representation of the effective octet and decuplet baryon fields
rather than the usual matrix expressions for the fields. There, to
one loop for equal mass light quarks in an expansion in $m_s$, we
found the appearance of all one- and two-body mass operators
listed in Eq. (\ref{hyperc-splittings}), but no three-body
operators.

For the case of nonzero $m_d-m_u$, it is straightforward to show
that one-loop corrections involving $M^d$ insertions and the
effects of meson mass differences in loop integrals introduce all
possible one- and two-body isospin splitting operators shown in
Eq. (\ref{group3}). We find no appearance of the three-body
isospin splitting operators at this level.

Recall that, in \cite{DH05}, our calculations of the
electromagnetic contributions to baryon masses were carried out
including the one-loop mesonic corrections to the basic
electromagnetic interactions, so to two loops overall. To this
order, the electromagnetic contributions also produce a complete
set of the one- and two-body mass splitting operators, but again
no three-body operators are generated. A general proof on the
cancellation of the three-body terms up to two loops will be given
in the next subsection.

As a result, a general parametrization of baryon masses to two
loops is
\beq \label{B-mass} \fl \mathcal{H}_{B} = m_0\bone +A \sum_{i\not
= j}\bsigma_i\cdot\bsigma_j + B \sum_iM^s_i + C \sum_{i\not
=j}M^s_i\bsigma_i\cdot\bsigma_j + D \sum_{i\not =j}M^s_iM^s_j +
\mathcal{H'}_{IB} \, , \eeq
where $\mathcal{H'}_{IB}$ is the isospin splitting mass
Hamiltonian consisting of the one- and two-body operators only,
\beq \label{Hib1} \fl \mathcal{H'}_{IB} = A' \sum_iM^d_i + B'
\sum_{i\not =j}M^d_i\bsigma_i\cdot\bsigma_j + C' \sum_{i\not
=j}M^d_iM^s_j + D' \sum_{i\not =j}M^d_iM^d_j \, . \eeq
Again, $m_0$, $A$, $B$, $C$, $D$, $A'$, $B'$, $C'$, and $D'$ are
the parameters.

Before moving on to the next subsection, we want to make clear
that, in this subsection, we are not {\it deriving} the
parametrization from the calculations, but are summarizing what
happens in actual dynamical calculations. We emphasize that the
form of the parametrization  above follows rigorously from
properties of QCD as shown by Morpurgo \cite{Morpurgo_param}, and
is encompassed also in the most general chiral Lagrangian in
effective field theory. While we obtained the same one- and
two-body operators in our earlier calculations in HBChPT
\cite{DHJ1,DHJ2,DH05} as summarized here, the form of the general
parametrization, whether approached from QCD or effective field
theory, is independent of the results of particular dynamical
calculations. The situation is different in the next subsection,
which deals specifically with the the origin of three-body effects
in HBChPT.

\subsection{Cancellation of the three-body terms}
\label{3bd-cancel}

\begin{figure}
\caption{\label{fig1} Time-ordered one- and two-loop three-body
diagrams which cancel exactly with renormalization diagrams. A
solid vertical line represents a quark moving upwards toward later
times. The horizontal dot-dashed lines pick out the intermediate
states. A squared dot indicates a one-particle or one-index
insertion. The zigzag line is a two-particle or two-index
insertion with no propagator or time ordering. A bold solid line
represents the exchange of a particle in the HBChPT Lagrangian
between the baryon and itself (a loop diagram in the baryon
picture). (a): One-loop exchange diagram with an one-particle
``mass" insertion. (b): The exchange diagrams with a two-particle
``mass" insertion where both time orderings of the particle
exchange vertices are shown. (c): A double exchange diagram with
the exchanges connecting the quark lines $i$ and $j$, and $j$ and
$k$. (d): A double exchange diagram similar to (c), but with an
one-particle insertion on one of the quark lines. Note that in
(a), (c), and (d), only one of the possible time orderings is
shown and in (b) both the possible orderings are included. There
are two, sixteen, and sixteen possible orderings for each type of
diagram in (a), (c), and (d), respectively.}
\begin{indented}
\item[]
\includegraphics{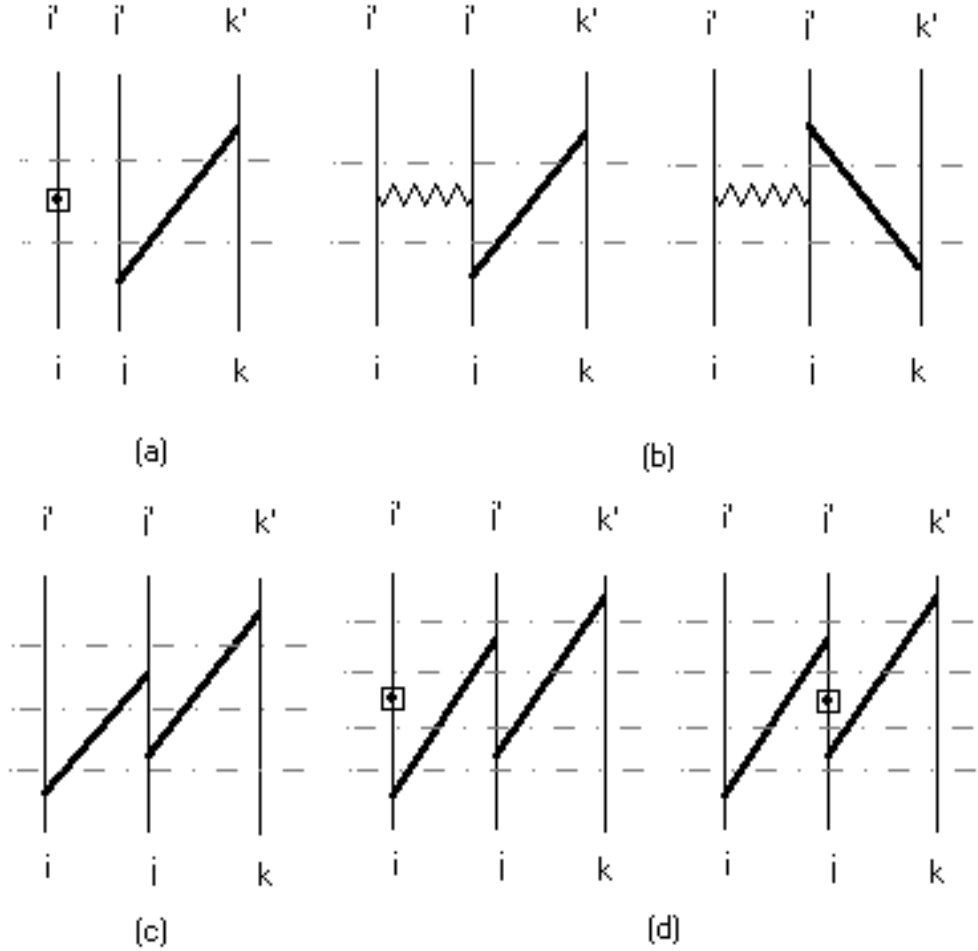}
\end{indented}
\end{figure}

We present here a general proof on the cancellation of the
three-body terms through two loops in HBChPT. Our analysis is done
using an expansion in meson (or photon) loops in time-ordered
perturbation theory. The four possible types of  time-ordered
diagram that involve all three indices $i,\,j,\,k$  (``three-body
operators'') and one- or two meson (or photon) loops are shown in
Fig. \ref{fig1}.  We are going to show that these diagrams cancel
exactly with the renormalization diagrams with the same topology.

In Fig. \ref{fig1}, a solid vertical line represents a ``quark" (a
flavor index) moving upward in time from an initial state with
flavor index $i$ to a final state with index $i'$ rather than a
propagating QCD quark. These flavor indices $i,j,k$ and $i',j',k'$
are connected in the effective field theory to the corresponding
indices on initial and final effective baryon fields
$\psi_{i,j,k}^\lambda$, $\psi_{i',j',k'}^{\lambda'}$ as described
in detail in \cite{DHJ1,DHJ2}, and the spin matrix elements are
calculated with respect to these fields. The baryon fields are
suppressed in the present graphical notation, and only the
connections of initial and final flavor indices and the structure
of the spin-dependent operators are shown explicitly.

A bold solid line connecting flavor lines represents the exchange
of a particle in the chiral effective theory such as a meson or
photon between flavor (or ``quark'') lines.  This exchange
corresponds in the baryon picture to a diagram with a meson or
photon loop connecting the baryon to itself, with the vertex
operators acting on different flavor and spin indices at different
times in time-dependent perturbation theory. The counting of loops
is done in this baryonic context. Note that the horizontal
dot-dashed lines pick out the intermediate states. The vertex
operators  typically involve flavor operators such as Gell-Mann
$\lambda$ matrices or charge or mass matrices, and spin operators
expressed in terms of Pauli matrices.

Similarly, a squared dot represents an operator or mass insertion
that acts only on one index, while a zigzag line represents a
``two-particle'' or two-index insertion with no propagator or time
ordering.

Note that in Figs. \ref{fig1}(a), \ref{fig1}(c), and
\ref{fig1}(d), only one of the possible time orderings is shown
and in Fig. \ref{fig1}(b) both the possible orderings are
included. There are two, sixteen, and again sixteen possible
orderings for each type of diagram in (a), (c), and (d),
respectively.

Firstly, let us consider Fig. \ref{fig1}(a) depicting a one-loop
exchange diagram with a one-particle ``mass" insertion. As will be
illustrated below, the value of the diagram can be expressed as a
product of  energy denominators and a matrix element of  vertex
operators dependent on the flavor and spin indices. In the figure,
the particle exchange connects the quark lines $j$ and $k$ and the
``mass" insertion is on the quark line $i$ at a time between two
vertices of the exchange line. The ``mass" insertion could include
a mass matrix, a spin, a charge matrix, a moment operator, and so
on; the latter two appeared in our analysis of baryon moments. The
matrix element involves all three indices $i$, $j$, $k$ so the
diagram is three-body in Morpurgo's classification. However, the
three vertices are on separate lines so they are not constrained
topologically. Moving the $j$ and $k$ vertices along those quark
lines to above or below the $i$ vertex, which can be done freely,
gives diagrams topologically equivalent to renormalization
diagrams, with the same vertex factors. In addition, the energy
denominator for the diagram corresponds to the one in the
renormalization constant associated with the exchange. Multiplying
the single particle mass insertion amplitude, which corresponds
diagrammatically to Fig. \ref{fig1}(a) with the particle exchange
deleted, by the renormalization constants for the initial and
final states associated with the particle exchange on the other
lines gives a term with the same energy denominator. This cancels
the three-body contribution from the diagram under consideration.

As a demonstration, we now show the time-ordered perturbation
theory calculations in HBChPT for the diagram in Fig.
\ref{fig1}(a) when the exchange is a meson. This corresponds in
the baryon picture to a one-loop diagram with the meson loop
connecting the baryon to itself. An energy denominator
$1/(E_0-E_n)$ appears in the expression for the perturbed energy
for each intermediate state $|n\rangle$ of the baryon system
between successive vertices. In the heavy-baryon approximation,
any internal baryon momenta $k$ resulting from baryon interactions
are small on the scale of the baryon mass $M$ and can be neglected
\cite{Jenkins1}.  The baryon energy is then always $M$ for the
initial baryon at rest, and cancels out in the difference
$E_0-E_n$. Hence, the energy factor reduces simply to
$-1/\sum_iE_i$, where the sum is over the energies of the lines
cut in the intermediate state, but with no contribution from the
quark lines. In this example, $E_i=E_l(\bk)=\sqrt{{\bk}^2+M_l^2}$,
where $M_l$ is the mass of the meson in the loop and $\bk$ is its
momentum in the baryon rest frame. We find that a contribution
from the diagram shown in Fig. \ref{fig1}(a) is
\beqarray \fl
\frac{\beta^2}{4f^2}\sum_l\int\frac{d^3k}{(2\pi)^3\,2\sqrt{{\bf
k}^2+ M_l^2}}\frac{F^2({\bk}^2)}{[-E_l({\bk})]^2}\,{\cal O}_{i'i}
\left(\lambda^l_{j'j} \,\bsigma_j\!\cdot\!{\bk} \,\bsigma_k
\!\cdot\! {\bk}\,\lambda^l_{k'k}\right) \nonumber \\
 = \frac{1}{3}\sum_l I'_l\, {\cal O}_{i'i}
\left(\lambda^l_{j'j}\, \bsigma_j \!\cdot\!
\bsigma_k\,\lambda^l_{k'k} \right) \, , \eeqarray
where ${\cal O}_{i'i}$ is the ``mass" insertion vertex, $\lambda$
is a Gell-Mann matrix in flavor space, $F(k^2)$ is a form factor
used to regularize the integral, and $I'_l$ is a modified integral
defined as
\begin{equation}
\label{I'_l} I'_l = \frac{\beta^2}{16\pi^2f^2}\int_0^\infty dk\,
\frac{k^4}{(k^2+M_l^2)^{3/2}}\, F^2(k^2) \, .
\end{equation}

A contribution from the renormalization diagram can be found by
multiplying the ``mass"-insertion amplitude ${\cal O}_{i'i}$ by
the wave function renormalization constants $Z=1-\delta Z$ for the
initial and final states associated with the meson exchange on
line $j$ and $k$. The result is
\beq -\frac{1}{2} \left({\cal O}_{i'i} \, \delta Z_{j'k';jk} +
\delta Z_{j'k';jk} \, {\cal O}_{i'i} \right) = - {\cal O}_{i'i} \,
\delta Z_{j'k';jk} \, , \eeq
where
\beq \label{deltaZ} \delta Z_{j'k';jk} = \frac{1}{3}\sum_l I'_l
\,\lambda^l_{j'j} \lambda^l_{k'k} \,\bsigma_j\!\cdot\!\bsigma_k \,
. \eeq
It is obvious that contributions from the diagram in Fig.
\ref{fig1}(a) and from its renormalization diagram cancel exactly
as suggested by its topology.

Secondly, we turn to the case of the two-particle ``mass"
insertion. Again, we consider the exchange diagrams with a
particle exchange connecting the quark lines $j$ and $k$, but with
a two-particle ``mass" insertion ``connecting" quark lines $i$ and
$j$ as shown in Fig. \ref{fig1}(b). Note that the two-particle
insertion, denoted by ${\cal O}_{i'i,j'j}$, is an instantaneous
short-distance term (e.g., ${\cal O}_{i'i,j'j}=(M^p \bsigma \cdot
\bsigma)_{i'i,j'j}$, where $p=d,s$) but with one-loop structure,
so we count it as equivalent to one loop. For the first (second)
ordering in Fig. \ref{fig1}(b), since ${\cal O}_{i'i,j'j}$ has no
effect on the vertex on line $k$, and its vertex on line $j$ is
above (below) the exchange vertex on that line, the exchange
vertices can be moved freely down (up) on lines $j$ and $k$ to
obtain a figure with the topology of a renormalization diagram. On
the other hand, because the two-particle insertion has no energy
denominator and does not contribute to $\delta Z$, the energy
denominators for both the orderings in Fig. \ref{fig1}(b) are
identical to those for the renormalization diagrams with the same
topology. As a result, when all contributions are added together,
the three-body terms from the exchange diagrams and their
renormalization diagrams cancel each other as expected. For
example, if the exchange particles are the mesons, the diagrams in
Fig. \ref{fig1}(b) give the following contribution
\beq \frac{1}{3}\sum_l I'_l\, \left( {\cal O}_{i'i, j'j''}
\lambda^l_{j''j}\, \bsigma_j \!\cdot\! \bsigma_k\,\lambda^l_{k'k}
+ \lambda^l_{j'j''}\, \bsigma_{j''} \!\cdot\!({\cal O}_{i'i, j''j}
 \, \bsigma_k\,\lambda^l_{k'k}) \right) \, ,  \eeq
that cancels exactly with the contribution from the
renormalization diagrams
\beq -\left({\cal O}_{i'i, j'j''} \, \delta Z_{j''k';jk} +  \delta
Z_{j'k';j''k} \, {\cal O}_{i'i, j''j} \right) \, , \eeq
where $\delta Z$ is given by Eq. (\ref{deltaZ}).

Thirdly, in Fig.  \ref{fig1}(c), we consider a double exchange
diagram with the exchanges connecting, say, $ij$ and $jk$. The
vertices on lines $i$ and $k$ can be freely slid along those
lines, while the two vertices on $j$ can be slid in opposite
directions. Moving the $ij$ exchange up leads to a figure
topologically equivalent to that encountered for an $ij$ exchange
multiplied by the renormalization constant for a $jk$ exchange, or
conversely. Recall that there are sixteen possible time orderings
for this type of diagram. When we combine the energy denominators
for various orderings, we in fact get those for products of single
exchange diagrams and renormalization factors. We also need to
include explicitly the contributions of single $ij$ and $jk$
exchanges multiplied by renormalization factors. The cancellation
occurs only when all contributions are added together.

As an example, we take the exchanges to be the mesons and find the
following contribution from all the time orderings of the double
exchange diagram
\beq \label{2loops} -\frac{2}{9}\sum_{l,l'} \, \left( I_l I'_{l'}+
I'_l I_{l'} \right) \,{\cal V}_{i'j'k'; ijk}
 \, , \eeq
where $I_l$ is the common integral for the exchange of meson $l$,
\begin{equation}
\label{I_l} I_l = \frac{\beta^2}{16\pi^2f^2}\int_0^\infty dk\,
\frac{k^4}{k^2+M_l^2}\,F^2(k^2),
\end{equation}
the modified integral $I'_l$ is given by Eq. (\ref{I'_l}), and the
group factor ${\cal V}_{i'j'k'; ijk}$ is
\beq {\cal V}_{i'j'k'; ijk}=(\bsigma_i \cdot \bsigma_j)(\bsigma_j
\cdot \bsigma_k) \, \left( \lambda^l_{i'i}(\lambda^l
\lambda^{l'})_{j'j}\lambda^{l'}_{k'k} +
\lambda^l_{i'i}(\lambda^{l'} \lambda^l)_{j'j}\lambda^{l'}_{k'k}
\right) \, . \eeq
To determine a contribution from the renormalization diagrams, the
amplitudes for the two separate $ij$ and $jk$ exchanges are
multiplied with the parts of the renormalization constant $Z$
associated with other exchanges. Sum of the two products gives
terms that cancel the original "three-body" amplitude shown in Eq.
(\ref{2loops}).

Fourthly, it is straightforward to generalize the above arguments
for a double exchange diagram with the exchanges connecting two
pairs of lines and an extra one-particle ``mass" insertion as
depicted in Fig. \ref{fig1}(d). Since we can always slide one
exchange line topologically outside the rest of the diagram,
whatever the initial time orderings of the vertices, the
cancellation occurs when all time orders are summed and all the
$\delta Z$'s associated with that line in the various time-orders
are included.

\begin{figure}
\caption{\label{fig2} Three distinct topologies of the three-loop
three-body diagrams that cannot be cancelled by renormalizations.}
\begin{indented}
\item[]
\includegraphics{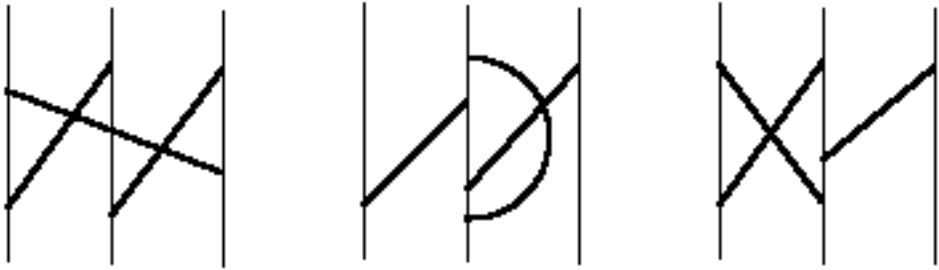}
\end{indented}
\end{figure}

Finally, we present in Fig.  \ref{fig2} the three-exchange
diagrams that cannot be cancelled by renormalizations. The same
three cases also appear when one of the exchanged lines is
replaced by an instantaneous two-particle ``mass'' insertion. One
can see that there are three distinct topologies involved:  i) two
vertices on each line with the loops entangled;  ii) one vertex on
each of two lines and four vertices on the remaining line with a
loop from that line to itself enclosing the other two vertices;
and iii) two vertices on each of two lines with crossed exchanges,
and a third trapped vertex on one of the lines with the exchange
connecting to the third line. In the remaining three-loop
diagrams, one line can always be slid outside the remainder of the
diagram, leading topologically to a renormalization-type diagram,
and the final result is actually cancelled by renormalization
terms.

We conclude this subsection by noting that since the three-body
contributions only appear at three loops, one can expect their
contributions to the baryon masses to be very small.

\section{Hierarchy and Fits} \label{hierarchy-and-fits}

\subsection{Hierarchy of sizes} \label{hierarchy}

In the case of hypercharge breaking, Morpurgo
\cite{Morpurgo_hierarchy,Morpurgo_QM2} argued on the basis of
ideas on the interactions of gluons with structure (rather than
elementary) quarks and fits to some of the coefficients that the
sizes of the coefficients of the various operators shown in the
most general expression for baryon masses (see Eqs.
(\ref{mass-gen}) - (\ref{Hib})) should satisfy a set of
hierarchical relations, with suppressions by a factor of $\sim
1/3$ for each factor of $M^s$ and by a factor in the range from
$0.22 - 0.37$ for each extra gluon exchange needed to get the
number of indices and spin factors in the operator
\cite{Morpurgo_EM2,DM-03} \footnote{Dillon and Morpurgo have
adopted the value of 0.3 for the suppression factor due to each
extra gluon exchange \cite{DM-03}.}.

Somewhat different arguments apply in the HBChPT approach when the
higher-order operators are generated by meson loop corrections to
a basic set, as in \cite{Jenkins1} and \cite{DHJ2}. Factors of
$M^s$  then appear multiplying differences of loop integrals
involving kaons or eta mesons and pions. The coefficients are
small because these differences would vanish for equal-mass
mesons.  $M^s$  also appears in explicit non-calculable mass
insertions with fairly small coefficients, smaller for
spin-dependent than spin-independent insertions.  These effects
were studied in detail in our previous work \cite{DHJ2}.

Similar results hold in the case of isospin breaking, with factors
of $M^d$ now multiplying either differences of pion (or kaon) loop
integrals for mesons within the isospin multiplet, or explicit
$d$-quark mass insertions.  Since $m_d<<m_s$, and the mass
differences within the pion and kaon multiplets are much smaller
than those between different multiplets, the coefficients of $M^d$
are expected to be much smaller than the corresponding
coefficients of $M^s$. The coefficients are also expected to be
smaller for spin-dependent than for spin-independent forms. The
results are consistent with the observed accuracy of various sum
rules for the masses.

It is important to note that the sizes of terms proportional to
$M^dM^d$, $M^dM^s$, or $M^sM^s$ do not go strictly as
$(m_d-m_u)^2$, $(m_d-m_u)(m_s-m_u)$, or $(m_s-m_u)^2$ as mass
insertion arguments would seem to suggest. The first two can both
be proportional instead in the loop expansion of HBChPT to $m_d -
m_u$  because of the structure of the loop corrections, while the
last can vary as $m_s-m_u$. This can be seen explicitly in the
$M^s \bsigma \cdot \bsigma$ and $M^sM^s \bsigma \cdot \bsigma$
terms obtained from the one-loop mesonic corrections to the baryon
masses in \cite{DHJ2}. The leading $M^s$-dependent corrections,
given in Eq. (3.14) in that paper, are all proportional to
differences  of meson loop integrals which vanish as $m_s-m_u$ as
the quark and meson masses become equal. A single factor of $M^s$
would therefore be expected. However, because of the particular
tensor structures involved in the products of Gell-Mann $\lambda$
matrices in the two-body couplings,  the loop corrections that
involve meson exchanges between quarks contribute to {\it both}
the $M^s \bsigma \cdot \bsigma$ and $M^sM^s \bsigma \cdot \bsigma$
terms, with similar sized corrections. The first is similar in
size to other contributions to the $M^s \bsigma \cdot \bsigma$
coefficient.  The second is much larger than would be expected
from the simple QCD argument.

Similar results hold for the meson loop corrections that involve
isospin breaking.  Since terms in $M^dM^d$ and $M^dM^s$ can be of
the same order as terms with a single factor of $M^d$, it is not
justified in applications to drop them without further analysis.
Thus, while contributions that explicitly involve higher powers of
the quark or meson mass differences, or higher loop integrals, are
clearly (or presumably) suppressed, we do not expect to find
Morpurgo's strict hierarchy for terms that involve multiple
factors of $M^d$ and $M^s$.

\subsection{Fits to the data} \label{fitdata}

Before doing the fits, we want to study the constraints on the
three-body terms. Note that without the three-body terms, we have
nine parameters $m_0$, $A$, $B$, $C$, $D$, $A'$, $B'$, $C'$, and
$D'$ to describe eighteen mass states of the baryon octet and
decuplet. Hence, we expect there to be nine sum rules among the
baryon masses. To find these sum rules, we construct a $18\times
18$ $\hat{\mathcal{M}}$ matrix  by replacing all columns
associated with the three-body operators in the $18\times 18$
$\mathcal{M}$ matrix defined earlier in Sect. \ref{gen-exp} with
zero entries. The obtained $\hat{\mathcal{M}}$ matrix has nine
zero eigenvalues. The sum rules are just the inner products of the
left null eigenvectors $\tilde{x}_0$ with
$\mu=\hat{\mathcal{M}}v$, $\tilde{x}_0\mu=
\tilde{x}_0\hat{\mathcal{M}}v=0$, where $v$ is a
eighteen-component column vector of coefficients $m_0$, $A$, $B$,
$C$, $D$, $a$, $b$, $c$, $A'$, $B'$, $C'$, $D'$, and $d_i$
$(i=1,..., 6)$. We find
\beqarray
\fl \Delta^0-\Delta^+ = n-p \nonumber \\
\fl \Delta^--\Delta^{++} = 3(n-p) \nonumber \\
\fl \Delta^0-\Delta^{++} = 2(n-p) + (\Sigma^0-\Sigma^+) - (\Sigma^--\Sigma^0) \nonumber \\
\label{sum_rules-ib}
\fl \Xi^--\Xi^0= (\Sigma^--\Sigma^+)-(n-p) \\
\fl \Xi^{*-}-\Xi^{*0} = (\Sigma^{*-}-\Sigma^{*+})-(n-p) \nonumber \\
\fl 2 \Sigma^{*0}-\Sigma^{*+} -\Sigma^{*-}=
2\Sigma^0-\Sigma^+-\Sigma^-  \nonumber \\
\fl \Xi^{*0}- \Xi^0 = \Sigma^{*+} -\Sigma^+ \nonumber \\
\fl \Omega^- -\Delta^{++}=3(\Xi^{*0}-\Sigma^{*+}) \nonumber \\
\fl (3p-n)/2+\Xi^0-(3\Lambda
+\Sigma^++\Sigma^0-\Sigma^-)/2=(3\Xi^{*0}-\Xi^{*-})/2-2\Sigma^{*+}+\Delta^{++}
\nonumber \, . \eeqarray
The first six relations are the well-known sum rules for isospin
splittings. The fourth is the Coleman-Glashow relation, suggested
originally on the basis of an unbroken SU(3) flavor symmetry
\cite{Coleman-Glashow} (another original SU(3) sum rule is the
Gell-Mann - Okubo (GMO) mass formula  \cite{Gell-Mann,Okubo}). All
the sum rules were later established for nonrelativistic quark
models with only one- and two-body interactions independently of
the flavor symmetry breaking
\cite{Rubenstein1,Rubenstein2,Ishida,Gal-Scheck}. Note that the
ninth relation is the modified GMO sum rule that first appeared in
the quark model context \cite{Rubenstein2}. If there are no
intramultiplet splittings, it reduces to the usual GMO rule.

The above sum rules are violated by three-body terms. If we
transfer all the terms to the left hand sides of the equations and
denote the differences by $\delta_i$ $(i=1, ..., 9)$, the results
for these sum rules in terms of the three-body coefficients are
\beqarray \fl \delta_1 &=& 6d_3 \, , \hspace{2.92 cm} \delta_2=
18d_3 + 6d_6 \, , \hspace{2.8 cm}
\delta_3= 6d_3 + 2d_4 - 4d_5 \, , \nonumber \\
\fl \delta_4 &=& 2d_1 -4d_2 - 2d_4 - 4d_5  \, , \, \, \delta_5=
2d_1 +2d_2 - 6d_3- 2d_4 - 2d_5 \, , \, \,
\delta_6= -6d_3 - 6d_5 \, , \label{delta} \\
\fl \delta_7 &=& -12a -6b \, , \hspace{1.9 cm} \delta_8= 6c \, ,
\hspace{1.6 cm} \delta_9= -6b + d_1 + d_2 +d_4 - 2d_5 \, .
\nonumber \eeqarray

{\bf \em Isospin splittings}: We have mentioned at the beginning
of this section that the coefficients of the mass operators in Eq.
(\ref{mass-gen}) are expected to satisfy an approximate hierarchy
of sizes. As discussed below, the sizes of some three-body
coefficients can be estimated by evaluating the $\delta$'s using
the experimental values of the accurately known baryon masses.

The $\Delta$ baryon masses are not determined with sufficient
accuracy for the first three sum rules to give a real test of this
expectation. The results for the next three as written are, in
order, $\delta_4=-0.31\pm 0.25\ \textrm{MeV}$, $\delta_5 = 0.09\pm
0.93\ \textrm{MeV}$, and $\delta_6 = -1.06\pm 1.18\ \textrm{MeV}$,
all consistent with zero within the experimental uncertainties. No
significant violations of the sum rules are evident.

Note that the coefficients $d_i$ $(i=1,..., 6)$ associated with
the three-body isospin splitting operators are expected to be very
small. In particular, the coefficient $d_6$ of the three-body
operator $t_6=\sum_{i\not=j\not=k}M^d_iM^d_jM^d_k$ that
contributes only to the $\Delta^-$ mass, is expected to be very
small since $t_6$ carries three factors of $M^d$ (its size is
proportional at least to $(m_d - m_u)^2$). We also expect that the
coefficients $d_5$ associated with the three-body operators
$t_5=\sum_{i\not=j\not=k}M^d_iM^d_jM^s_k \bsigma_i\cdot\bsigma_k$
should be smaller than the other $d_i$ $(i=1,..., 4)$ since $t_5$
involves two factors of $M^d$, one factor of $M^s$, and a spin
interaction with the third particle. Therefore, we will ignore
$d_5$ and $d_6$ when trying to estimate the other coefficients
$d_i$. Using the values of $\delta_4$, $\delta_5$, and $\delta_6$,
we find $d_1-d_4=0.33 \pm 0.50$ MeV, $d_2=0.24 \pm 0.25$ MeV, and
$d_3=0.18 \pm 0.20$ MeV. These results show that contributions of
the three-body isospin splitting operators to baryon masses are
indeed very small and can be ignored.

{\bf \em Hypercharge splittings}: If we set $d_i=0$ $(i=1,...,
6)$, as the above results suggest, we are left with the relations
\beq \label{delta789} \delta_7 = -12a -6b \, , \, \, \delta_8= 6c
\, , \, \, \delta_9= -6b \, . \eeq
To estimate the coefficients of the three-body hypercharge
splitting terms $a$, $b$, and $c$, we use the average mass of the
$\Delta$'s and other accurately known baryon masses to evaluate
$\delta_7$, $\delta_8-\delta_9$ (note that $\Delta^{++}$ does not
appear in this combination), and $\delta_9$. The result are
$\delta_7 = -23.54 \pm 0.55 \ \textrm{MeV}$,
$\delta_8-\delta_9=3.40 \pm 1.08 \ \textrm{MeV}$, and $\delta_9 =
-9.95 \pm 2.24 \ \textrm{MeV}$. Then, Eq. (\ref{delta789}) gives
$a=1.13 \pm 0.06$ MeV, $b=1.66 \pm 0.37$ MeV, and $c=-1.09 \pm
0.41$ MeV. These contributions of the three-body hypercharge
splitting terms are statistically significant and as large as
observed isospin splittings, so they should be kept in a complete
analysis of the baryon octet and decuplet masses.

{\bf \em General constrained fit}: We now consider a fit to the
baryon masses using the general expression in Eq. (\ref{mass-gen})
neglecting the coefficients $d_i$ $(i=1,..., 6)$. Using 12
parameters $m_0$, $A$, $B$, $C$, $D$, $A'$, $B'$, $C'$, $D'$, $a$,
$b$, and $c$, we can do a weighted least-squares fit to the 15
measured quantities (14 accurately known baryon masses plus the
average mass of the $\Delta$'s).

\begin{table}
\caption{Baryon masses in units of MeV. Excluding the $\Delta$
resonances, the average deviation $\overline{|\Delta M_B|} = 0.07
$ MeV, where $\Delta M_B = M_B^{\textrm{theory}}-
M_B^{\textrm{expt.}}$. The experimental data are from the listings
given by the Particle Data Group \cite{PDG}. No definite masses or
uncertainties are given for the $\Delta$ resonances as different
experimental analyses are in conflict. Only ranges of mass are
given here.}
\begin{indented}
\item[]
\begin{tabular}{@{}lccc}
\br
Baryon & Theory & Expt. & $|\Delta M_B|$ \\
\mr
$p$ & 938.27 $\pm$ 1.50 & 938.27 $\pm$ 0.00 & 0.00  \\
$n$ & 939.57 $\pm$ 1.51 & 939.57 $\pm$ 0.00 & 0.00 \\
$\Lambda$ & 1115.68 $\pm$ 2.10 & 1115.68 $\pm$ 0.01 & 0.00  \\
$\Sigma^+$ & 1189.39 $\pm$ 2.28 & 1189.37 $\pm$ 0.06 & 0.02 \\
$\Sigma^0$ & 1192.64 $\pm$ 2.30 & 1192.64 $\pm$ 0.02 & 0.00 \\
$\Sigma^-$ & 1197.45 $\pm$ 2.36 & 1197.45 $\pm$ 0.03 & 0.00 \\
$\Xi^0$ & 1314.64 $\pm$ 3.58 & 1314.83 $\pm$ 0.20 & 0.19 \\
$\Xi^-$ & 1321.39 $\pm$ 3.62 & 1321.31 $\pm$ 0.13 & 0.08 \\
$\Delta$ & 1232.00 $\pm$ 1.48 & 1232.00 $\pm$ 2.00 & 0.00  \\
$\Delta^{++}$ & 1230.82 $\pm$ 1.46 & 1230.5 $\pm$ 0.2 - 1231.88 $\pm$ 0.29 & $\cdots$ \\
$\Delta^+$ & 1230.57 $\pm$ 1.47 & 1231.6 - 1234.9 $\pm$ 1.4 & $\cdots$ \\
$\Delta^0$ & 1231.87 $\pm$ 1.53 & 1233.1 $\pm$ 0.3 - 1234.35 $\pm$ 0.75 & $\cdots$ \\
$\Delta^-$ & 1234.73 $\pm$ 1.64 & $\cdots$ & $\cdots$ \\
$\Sigma^{*+}$ & 1382.74 $\pm$ 1.95 & 1382.80 $\pm$ 0.40 & 0.06 \\
$\Sigma^{*0}$ & 1384.18 $\pm$ 2.02 &  1383.70 $\pm$ 1.00 & 0.48 \\
$\Sigma^{*-}$ & 1387.18 $\pm$ 2.06 & 1387.20 $\pm$ 0.50 & 0.02 \\
$\Xi^{*0}$ & 1531.81 $\pm$ 3.37 & 1531.80 $\pm$ 0.32 & 0.01 \\
$\Xi^{*-}$ & 1534.95 $\pm$ 3.41 & 1535.00 $\pm$ 0.60 & 0.05 \\
$\Omega^-$ & 1672.45 $\pm$ 6.78 & 1672.45 $\pm$ 0.29  & 0.00 \\
\br
\end{tabular}
\label{massvalues}
\end{indented}
\end{table}
\normalsize
A best fit is obtained at values (in MeV)
\beqarray \label{bestfit1} \fl m_0 = 1082.9 \pm 1.0 \, ,
\nonumber \\
\fl A=24.66 \pm 0.17 \, , \, \, B= 183.66 \pm 1.05 \, , \, \,
C=-17.10 \pm 0.34 \, , \, \, D=-2.94 \pm 0.71 \, , \nonumber
\\
\fl a=1.22 \pm 0.17 \, , \, \,  b=1.52 \pm 0.35 \, , \, \,  c=-0.93 \pm 0.40 \, , \nonumber \\
\fl A'= 0.95 \pm 0.18 \, , \, \,  B'=-0.60 \pm 0.08 \, , \, \,
C'=1.69 \pm 0.24 \, , \, \, D'=0.78 \pm 0.04  \, . \eeqarray
The fitted masses have an average deviation from experiment of
only 0.07 MeV and a $\chi^2=1.73$ (with 3 degrees of freedom). The
calculated values of the baryon masses are given in Table
\ref{massvalues} where the experimental data are from the listings
given by the Particle Data Group \cite{PDG}. No definite masses or
uncertainties are given for the $\Delta$ resonances as different
experimental analyses are in conflict. Instead, we give here their
mass ranges except for $\Delta^-$. Our predictions for the
$\Delta$ resonances agree with the experimental values within the
margin of error.

Note that the best-fit values of the parameters $A'$, $B'$, $C'$,
and $D'$ show that terms in $M^dM^d$ and $M^dM^s$ are of the same
order as terms with a single factor of $M^d$, as expected in the
meson loop expansion, rather than much smaller as suggested by the
QCD hierarchy arguments.

{\bf \em Baryon mass differences}: We remark that one can use the
general expression for baryon masses in Eq. (\ref{mass-gen}) to
describe the baryon mass differences. Then the parameters
associated with two flavor-symmetric operators and with the
intermultiplet splitting operators drop out, so there remain 10
parameters ($A'$, $B'$, $C'$, $D'$, and $d_i$ $(i=1,..., 6)$).
Using contributions of the isospin splitting operators to the mass
differences given in Table \ref{IB-terms}, we easily find the
followings
\begin{eqnarray} \label{splittings}
n - p & = & A' + 2B'+ 2D'-4d_3 \, , \nonumber \\
\Sigma^- - \Sigma^+ & = & 2(A'-B'+C'+D'-2d_3+d_4-2d_5) \, , \nonumber \\
\Sigma^- - \Sigma^0 & = & A'-B'+C'+2D'-4d_3+2d_4-4d_5 \, , \nonumber \\
\Xi^- - \Xi^0 & = & A'-4B'+2C'+2d_1-4d_2
\, , \nonumber \\
\Sigma^{*-} - \Sigma^{*+} & = & 2(A'+2B'+C'+D'+d_3+d_4+d_5) \, , \\
\Sigma^{*-} - \Sigma^{*0} & = & A'+2B'+C'+2D'+2d_3+2d_4+2d_5 \, , \nonumber  \\
\Xi^{*-} - \Xi^{*0} & = & A'+2B'+2C'+2d_1+2d_2 \, , \nonumber \\
\Delta^0 - \Delta^{++} & = & 2(A'+2B'+D'+d_3) \, , \nonumber \\
\Delta^- - \Delta^{++} & = & 3(A'+2B'+2D'+2d_3+2d_6) \, , \nonumber \\
\Delta^0 - \Delta^+ & = & A'+2B'+2D'+2d_3 \, . \nonumber
\end{eqnarray}

If the three-body isospin splitting terms $d_i$ are neglected, we
have 4 parameters $A'$, $B'$, $C'$, $D'$ to describe the baryon
mass differences. We studied this case earlier in \cite{DH05,Ha07}
using the one- and two-body electromagnetic operators. A best
weighted fit to the seven known mass splittings other than those
for the $\Delta$ baryons is obtained at values (in MeV) of $A' =
0.91 \pm 0.19$, $B' = -0.59 \pm 0.08$, $C' =1.73 \pm 0.23$, and
$D' = 0.78 \pm 0.05$ with an average deviation from experiment of
0.13 MeV and a $\chi^2=1.67$ (with 3 degrees of freedom). The
calculated values of the baryon mass splittings are the same as
those shown in Table I in \cite{Ha07}.

Our above fits provide the parameters that must be explained in
the loop expansion in chiral effective field theory or any other
dynamical model, and show that three-body contributions will
necessarily play a role in explaining masses at the scale of about
1 MeV. The scale for three-body isospin effects is at least ten
times smaller.

\section{Conclusions} \label{conclusions}

We have constructed the general parametrization of the baryon
octet and decuplet mass operators including three-body terms using
the unit operator and the symmetry-breaking factors, $M^d$ and
$M^s$, in conjunction with the spin operators. Our general
expression for baryon masses includes all order of the light quark
mass difference $m_d-m_u$. Using the Gram matrix analysis, we
established a minimal set of 18 independent operators to describe
18 baryon octet and decuplet mass states.

We have demonstrated the cancellation of the three-body terms
through two loops in HBChPT and have identified three distinct
topologies of the three-loop diagrams that generate the three-body
terms. We also investigated the likely size of the three-body
terms using the sum rules that give constraints on their
coefficients. Numerical calculations of the sum rules indicate
that contributions of the three-body hypercharge splittings, while
small, are comparable to those from the one- and two-body isospin
splittings. The contributions of three-body isospin splitting
operators are statistically consistent with zero, and these
operators can be neglected at the present level of experimental
accuracy. We find that the suggested hierarchy of sizes for terms
that involve multiple factors of $M^d$ and/or $M^s$ may not be
evident because powers of $M^d$ and $M^s$ do not correlate
strictly with powers of the quark masses $m_d-m_u$ and $m_s-m_u$,
respectively.

We have done the phenomenological fits to the experimental data
using our general expression, but ignoring terms associated with
the three-body isopsin splitting operators. Our fits provide the
values of the parameters that must be explained in the loop
expansion in ChPT or any other dynamical model. The results of the
fits also confirm that three-loop contributions will necessarily
play a role in explaining masses at the scale of about 1 MeV.

Finally, our parametrization has the minimal number of operators
needed to describe all the octet and decuplet masses. It can be
translated back into the language of the heavy-baryon effective
field theory and is particularly useful to an analysis of the
baryon mass splittings due to both hypercharge-breaking and
isospin-breaking effects.

\ack

The author would like to thank Professor Loyal Durand for useful
comments and invaluable support.


\end{document}